\documentclass{article}

\newcommand {\apj}{Astrophys. J.}
\newcommand{\aap}{Astron. Astrophys.}
\newcommand{\prl}{Phys. Rev. Lett.}
\newcommand{\prd}{Phys. Rev. D}
\newcommand{\prc}{Phys. Rev. C}

\newcommand{\plb}{Phys. Lett. B}
\usepackage{psfig, epsfig,rotate,multicol}
\begin{document}

\title{Bulk viscosity of  strange quark matter in density-dependent quark mass model and dissipation of r-modes in strange stars }
\author{Zheng Xiaoping, Liu Xuewen, Kang Miao, Yang Shuhua
\\
{\small  Department of Physics, Huazhong Normal University,
Wuhan430079 P. R. China }}
\date{}
\maketitle
\begin{abstract}
We study the bulk viscosity of strange quark matter(SQM) in
density dependent quark mass model(DDQM) under the background of
self-consistent thermodynamics. The correct formulae, with which
the viscosity can be evaluated, are derived. We also find that the
viscosity in DDQM can be higher by 2 to 3 orders of magnitude than
MIT bag model. We calculate the damping time scales due to
viscosity coupled to r-modes. The numerical results show the time
scale can't be shorter than $10^{-1}$s.
\end{abstract}
\section{INTRODUCTION}

Since Witten conjectured\cite{1} that the strange quark
matter(SQM) composed of comparable number of $u$, $d$ and $s$
quarks might be absolutely stable or metastable phase of nuclear
matter, many theoretical and observational efforts have been made
on the investigation of its properties and potential astrophysical
significance\cite{2,3,4,5,6,7}. Because the difficulty of  quantum
chromodynamics (the lattice approach) in the non-perturbation
domain , ones used to adopt phenomenological models. One of the
famous models is the MIT bag model with which many authors study
the equation of state(EOS) of SQM and the mass-radius relation of
the assumed strange stars\cite{3,4,7}. Another alternative model
is  density-dependent quark mass model(DDQM). Originally,  some
pioneer investigations\cite{8,9,10,11,12}  were carried out but
the wrong thermodynamics treatment is applied. Latterly, some
works dealt with the thermodynamics with density-dependent
particle masses self-consistently with which the structural
properties of strange stars have been re-discussed\cite{13,14}.

It had also been recognized the effects of the dynamics of SQM on
strange star's rotations by the time that the EOS of SQM was
gradually uncovered. Wang and Lu find that the non-leptonic weak
reaction  dominates  bulk viscosity of SQM\cite{15}. Further
investigations of the bulk viscosity also have been carried out by
many other authors in MIT bag model and the fruitful results have
been obtained\cite{16,17,18}. A recent study incorporate the
effective quark masses in quasiparticle description into the bulk
viscosity\cite{19} and have a significant result for relevant
astrophysics\cite{20}. In parallel, a previous work  made the
study of the bulk viscosity in mass-density-dependent model but
the self-consistent thermodynamics  are overlooked\cite{21}. Here
we will concentrate
 the problem in the background of the correct thermodynamics and
 consider the effect of variable quark masses on the bulk viscosity of SQM.
As seen latter in this paper, the effect have twofold
contributions. The first, the modification of EOS influences the
viscosity. The second, the density-dependent masses directly
contribute to the dynamical quantity.

We organize this paper as such. In Sec.II, the self-consistent
thermodynamics with density-dependent particle masses are recalled
and then further the thermodynamical relations are discussed. In
Sec.III, the bulk viscosity of SQM with density-dependent particle
masses is derived, which arises from the non-leptonic weak
interaction. In Sec.IV, the numerical results  are given and a
 application to the recent astrophysical problem is discussed. In
 Sec. V, a brief summary is made.
 \section{THERMODYNAMICS WITH DENSITY\\
 -DEPENDENT PARTICLE MASSES}

 The self-consistent thermodynamics in mass-density-dependent
 model have been discussed in many works\cite{12,13,14}. However
 Peng et al\cite{14} pointed that the thermodynamics treatments given
 by\cite{9,10,11,12} have serious problem. They re-derived the pressure
 and energy expressions from the general ensemble theory and found
 an extra term can add to the expression of pressure but not to
 the  energy formula. Here we don't repeat the derivation but
 simply write the results. When particle masses depend on baryon number density or the volume of the system,
 the pressure and energy have
 \begin{equation}
 P=-\left. {\partial  (V\Omega(V))\over\partial V}\right |_{T,\{\mu\}},
 \end{equation}
\begin{equation}
 E=- \beta{\partial  (\beta V\Omega(V))\over\partial \beta}+\mu N+\beta{\partial (V\Omega)\over\partial\beta}.
 \end{equation}
  Although the thermodynamics potential density depends on the volume,
  we still obtain the thermodynamic relation
 \begin{equation}
E=V\Omega(V)+\mu N +TS,
 \end{equation}
where $\beta$ is the reverse temperature, $V$ is the system
volume, $\Omega$ is thermodynamics potential density, $\mu$ is the
chemical potential and $N$ is the particle number. According to
Eq.(3), the basic thermodynamic equation is written as at
invariable temperature.

 \begin{equation}
{\rm d}E={\rm d}(V\Omega(V))+{\rm d}(\mu
N)=(\Omega+V{\partial\Omega\over\partial V}){\rm d}V+\mu {\rm d}N,
 \end{equation}
Therefore, the pressure have an extra term
\begin{equation}
P=-\Omega-V{\partial\Omega\over\partial V},
\end{equation}
Equally, it is the formula
,$P=-\Omega+n_b{\partial\Omega\over\partial n_b}$, given in
\cite{14}. Those previous works have shown that the thermodynamics
 are self consistent, where  an extra term is appended to the
 pressure but not to the energy.
 In the framework of self consistent thermodynamics, the EOS of
 SQM and the structural properties of strange stars have been
 investigated. At zero temperature, the thermodynamic potential density is
 \begin{equation}
 \Omega=-\sum\limits_i{1\over 8\pi^2}\left [\mu_i(\mu_i^2-m_i^2)^{1/2}(2\mu_i^2-5m_i^2)
 +3m_i^4\ln{\mu_i+\sqrt{\mu_i^2-m_i^2}\over m_i}\right ].
 \end{equation}
 Accordingly, the energy and pressure  can be obtained from Eqs (3) and (5), and the number density is
 \begin{equation}
 n_i={1\over\pi}(\mu_i^2-m_i^2)^{3/2}.
 \end{equation}
  It has been found that the density-dependent quark
 masses should take the form\cite{13}
 \begin{equation}
 m_i=m_{i0}+{D\over n^{1/3}_b},\ \ \ \ i=u,d,s,
 \end{equation}
 where $D$ is a parameter to be determined by stability arguments.
 For given values $n_b$,
 \begin{equation}
 n_b={n_u+n_d+n_s\over 3},
 \end{equation}
 we very easily determine the quark chemical potentials $\mu_i$ to evaluate
 the thermodynamic quantities of the system by the beta equilibrium and the charge neutrality conditions
 \begin{equation}
 \mu_d=\mu_s\equiv\mu, \ \ \ \ \mu_s=\mu_u+\mu_e,
 \end{equation}
 \begin{equation}
 {2\over 3}n_u-{1\over 3}n_d-{1\over 3}n_s-n_e=0.
 \end{equation}

 In the following, we will consider the dynamics of SQM( bulk viscosity) near the
 equilibrium state, which is governed by the EOS calculated with
 the self consistent thermodynamics.
\section{DERIVATION OF THE BULK VISCOSITY}
 In a previou study\cite{21}, the formulae in MIT model, which is presented by Wang and Lu\cite{15}
  and developed by other investigators\cite{16,17,18},
   was simply introduced to evaluated the  bulk viscosity of SQM in density dependent quark mass model.
   The investigation used an incorrect thermodynamical results
   without regard to the consequences due to the dependence of the quark masses on baryon number density.
In this section, we derive the bulk viscosity of SQM in the self
 consistent thermodynamics. Since the quark masses are the
 functions of the baryon number density represented in(6),  we  for convenience
 assume the baryon number density instead of the volume as had adopted in
 previous works\cite{15,16,17,21}
 oscillates periodically,
 \begin{equation}
 n_b=n_{b0}+\Delta n_b \sin({2\pi t\over \tau})=n_{b0}+\delta n_b,
 \end{equation}
 where $n_{b0}$ is the equilibrium baryon number density and
 $\Delta n_b$ is the amplitude of perturbation, $\tau$ is the
 period. The presence of viscosity results in the energy dissipation of pulsations in bulk matter. On one hand,using
 the standard definition of the bulk viscosity of matter, energy dissipation rate per unit volume average
 over the oscillation period $\tau$ can be written as
 \begin{equation}
\dot{E}_{\rm kin}=-{\zeta\over\tau}\int^\tau_0{\rm d}t({\rm div}{\bf v})^2.
\end{equation}
where ${\bf v}$ is velocity of the fluid. On the other hand, the
hydrodynamic matter deviates from the equilibrium accompanied by
the time variations of the local pressure, $P(t)$. The dissipation
of the energy of hydrodynamic motion due to irreversiblity of
periodic compression-decompression process can be expressed as
\begin{equation}
\dot{E}_{\rm diss}=-{1\over\tau V}\int_0^{\tau}{\rm d}tP(t)\dot{V}.
\end{equation}
According to the continuity equation $\dot{n}_b+n_{b0}{\rm
div}{\bf v}$ and the specific volume $V={1\over n_b}$, bearing in
mind that $\dot{E}_{\rm kin}=-\dot{E}_{\rm diss}$, from Eqs(13)and
(14) we have the bulk viscosity
\begin{equation}
\zeta={1\over\pi}({n_{b0}\over\Delta n_b})\int_0^{\tau}P(t)\cos{2\pi t\over\tau}.
\end{equation}
  It is proven that the bulk viscosity of SQM is mainly produced by the nonleptonic interaction as follows
 \begin{equation}
 u+d\longleftrightarrow s+u
 \end{equation}
 thereby  at quasi-equilibrium the pressure can be regarded as a function of
the number density $n_i$, where $i$ takes $u,d,s$. From the baryon number density(9),
it is suitable to assume that $P=P(n_b, n_d, n_s)$. Thus the pressure can expand in small deviations
\begin{equation}
P=P_0+({\partial P\over\partial n_b})\delta n_b+({\partial P\over\partial n_d})\delta n_d+({\partial P\over\partial n_s})\delta n_s,
\end{equation}
where all derivatives are taken at equilibrium and the changes in d- and s-quark number density will be composed of two parts
\begin{equation}
\delta n_i=\delta\tilde{n}_i-n_i{\delta V\over V},
\end{equation}
where $\delta\tilde{n}_i$ denotes the number density at a given volume for s- and d-quark. From the reaction (16),
it evidently satisfies the relation below
\begin{equation}
\delta\tilde{ n}_d=-\delta\tilde{n}_s=\int_0^t{{\rm d}\tilde{n}_d\over{\rm d}t}{\rm d}t.
\end{equation}
 Due to the baryon number conservation, we also have
\begin{equation}
{\delta n_b\over n_b}=-{\delta V\over V}.
\end{equation}
Finally, Eq(17) becomes
\begin{equation}
P=P_0+\left [n_b{\partial P\over\partial n_b}+n_d{\partial P\over\partial n_d}+n_s{\partial P\over\partial n_s}\right ]{\delta n_b\over n_b}
+\left [{\partial P\over\partial n_d}-{\partial P\over\partial n_s}\right ]\int_0^t{{\rm d}\tilde{n}_d\over{\rm d}t}{\rm d}t.
\end{equation}
Obviously, only the latest term contributes to the integral in Eq(15). In the formula above the net rate per unit volume
can be transform  as from the rate per mass given by\cite{21}
\begin{equation}
{{\rm d}\tilde{n}_d\over{\rm d}t}={3\over\pi^3}G_F^2\sin^2\theta\cos^2\theta JT^2\left (1+{\delta\mu^2\over 4\pi^2T^2}\right )\delta\mu,
\end{equation}
where $J$ is given in \cite{22} and $\delta\mu=\mu_s-\mu_d$. Analogously to the pressure, the chemical potential difference can be expanded, giving
\begin{equation}
\delta\mu (t)=({\partial\delta\mu\over\partial n_b})\delta n_b+({\partial\delta\mu\over\partial n_d})\delta n_d+({\partial\delta\mu\over\partial n_s})\delta n_s.
\end{equation}
To obtain the reaction rate and then calculate the bulk viscosity, the acquire  of the derivatives ${\partial P\over\partial n_i}$ and
${\partial\mu_i\over\partial n_i}$ is necessary with the equation of state in section above. We respectively
regard the thermodynamic potential
and the chemical potential as the functional form below
\begin{equation}
\Omega=\Omega(\mu_j(n_i,n_b(n_i)), n_b(n_i)),\ \ \ \
\mu_i=\mu_i(n_i, n_b(n_j))
\end{equation}
where $i,j$ take $u,d,s$. From (5) and (7), we immediately obtain
\begin{equation}
{\partial P\over\partial n_j}=\sum\limits_i(n_i-3n_b){\partial\mu_i\over\partial n_j},
\end{equation}
\begin{equation}
{\partial\mu_i\over\partial n_i}={\mu_i^2-m_i^2\over 3\mu_i n_i},\ \ \ \
{\partial\mu_i\over\partial n_b}={m_i\over\mu_i}{\partial m_i\over\partial n_b}.
\end{equation}
Thus, the part in(21) contributed to the integral (15), $\delta P$, and the chemical potential difference at quasi-equilibrium, $\delta \mu$
can be expressed as
\begin{equation}
\delta P=\left ({n_d-3n_b\over n_d}{\mu^2-m_d^2\over 3\mu}-{n_s-3n_b\over n_s}{\mu^2-m_s^2\over 3\mu}\right )\int_0^t{{\rm d}\tilde{n}_d\over{\rm d}t},
\end{equation}
\begin{equation}
\delta\mu=-\left ({n_b(m_s-m_d)\over\mu}{\partial m_d\over\partial n_b}-{m_s-m_d\over 3\mu}\right ){\delta n_b\over n_b}
-\left ({\mu^2-m_d^2\over 3\mu n_d}+{\mu^2-m_s^2\over 3\mu n_s}\right )\int_0^t{{\rm d}\tilde{n}_d\over{\rm d}t}.
\end{equation}
Substituting Eq(27) into Eq(15), we finally get the viscosity
\begin{equation}
\zeta={1\over\pi}{n_{b0}\over\Delta n_b}\left ({n_d-3n_b\over n_d}{\mu^2-m_d^2\over 3\mu}-{n_s-3n_b\over n_s}{\mu^2-m_s^2\over 3\mu}\right )\int_0^\tau{\rm d}t\int_0^t{{\rm d}\tilde{n}_d\over{\rm d}t}.
\end{equation}
For the sake of convenience, Eq(28) can be  equivalently rewritten as
\begin{equation}
{\partial\delta\mu\over\partial t}=-\left ({n_b(m_s-m_d)\over\mu}{\partial m_d\over\partial n_b}-{m_s-m_d\over 3\mu}\right ){\delta n_b\over n_b}
-\left ({\mu^2-m_d^2\over 3\mu n_d}+{\mu^2-m_s^2\over 3\mu n_s}\right ){{\rm d}\tilde{n}_d\over{\rm d}t}.
\end{equation}
\section{RESULTS AND DISCUSSION}

Eqs (22) and (30) must be solved numerically to give the
rate${{\rm d}\tilde{n}_d\over{\rm d}t}$ and the numerical result
of Eq(29) can be obtained.

\begin{figure}[h]
\centerline{\epsfig{file=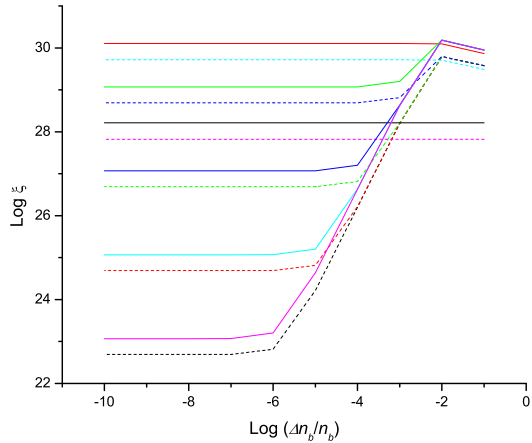,  width=8cm}} \caption{Bulk
viscosity as a function of relative perturbation ${\Delta n_b\over
n_b} $. Solid curves for $n_b=0.4$fm$^{-3}$, $D^{1\over
2}=156$MeV, $m_{s0}=80$MeV    and dashed curves for
$n_b=1.36$fm$^{-3}$, $D^{1\over 2}=156$MeV, $m_{s0}=140$MeV. The
both for $\tau=10^{-3}$ and temperature $10^{-5}, 10^{-4},
10^{-3}, 10^0, 10^{-2}, 10^{-1}$MeV from bottom to
top.}\label{mfig1}
\end{figure}

In figure\ref{mfig1} we have plotted viscosity vs. relative
perturbation ${\Delta n_b\over n_b} $ respectively for
(i)$D^{1\over 2}=156$MeV, $m_{s0}=80$MeV  at $n_b=0.4$fm$^{-3}$
and $\tau=10^{-3}$(solid curves) and (ii)$D^{1\over 2}=156$MeV,
$m_{s0}=140$MeV  at $n_b=1.36$fm$^{-3}$ and $\tau=10^{-3}$(dashed
curves).

\begin{figure}[h]
\centerline{\epsfig{file=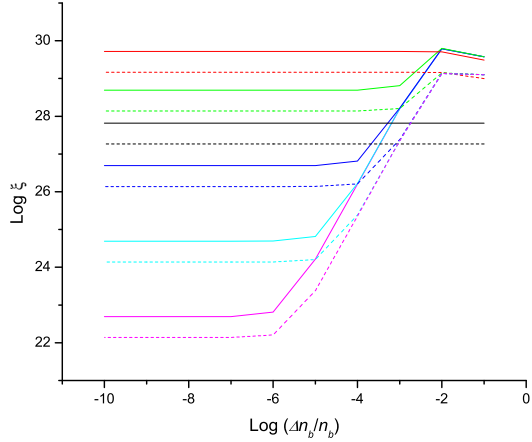, width=8cm,}} \caption{A
comparison of the our bulk viscosity(solid curves) with that in
ref[21](dashed curves)for  parameters $n_b=0.4{\rm
fm}^{-3},D^{1\over 2}=156.0{\rm Mev}, m_{s0}=80{\rm Mev}$
and$\tau=10^{-3}$. The corresponding temperatures of the curves as
figure \ref{mfig1}}\label{mfig2}
\end{figure}

In figure\ref{mfig2} we gave the results from our formulae(solid
curves) and those in ref.\cite{21}(dashed curves) for  the same
parameters $n_b=0.4{\rm fm}^{-3},D^{1\over 2}=156.0{\rm Mev},
m_{s0}=80{\rm Mev}$.  Evidently our viscosity can be higher.

\begin{figure}[h]
\centerline{\epsfig{file=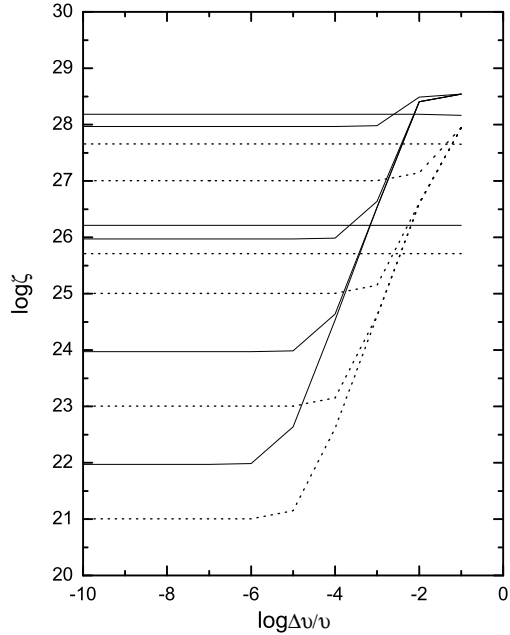, width=8cm}} \caption{Bulk
viscosity in MIT bag for $m_s$=80MeV , $\mu_d$=470MeV,
$\tau=10^{-3}$s. Solid curves for interacting quark matter, dashed
curves for ideal quark gas and the corresponding temperatures of
the curves as figure \ref{mfig1}}\label{mfig3}
\end{figure}

In figure\ref{mfig3} we show the results in MIT bag model from
Zheng et al\cite{19} for $m_{s0}=80{\rm MeV}, \tau=10^{-3}$. The
solid curves are because of the consideration of screening mass
effect and the dashed curves are regardless of the coupling among
quarks in MIT bag model. Comparing figure\ref{mfig2} with solid
curves in figure\ref{mfig3}, we notice that our results are higher
by 2 to 3 order of magnitude than those of bag model. We also find
that the solid curves in figure\ref{mfig2} is also higher than the
dashed curves. In a word, the facts testify to the viscosity
increase of SQM due to medium effect.

 Recently it
has been realized that rotating relativistic stars are generically
unstable against the r(otational)-mode instability\cite{23} The
viscous dissipation of dense matter has relevance to the onset of
r-mode instability, gravitational radiation detection and the
evolution of pulsars\cite{24,25,26,27}. A main motivation for the
studies of viscosity is to calculate the damping time for compact
star vibrations. For a star of constant density, the r-mode energy
is estimated as
\begin{equation}
E={\alpha^2\pi\over 2m(2m+3)}(m+1)^3(2m+1)!\rho R^5\Omega^2.
\end{equation}
The mode dissipation energy can be computed in term of
\begin{equation}
{{\rm d}E\over{\rm d}t}=\int\zeta\delta\sigma\delta\sigma^*{\rm d}^3x.
\end{equation}
 The damping time scale is thus
\begin{equation}
\tau_D=-{2E\over{{\rm d}E\over{\rm d}t}}=1.31\times
10^{15}\rho^2\Omega^{-4}\zeta^{-1}.
\end{equation}

\begin{figure}[h]
\centerline{\epsfig{file=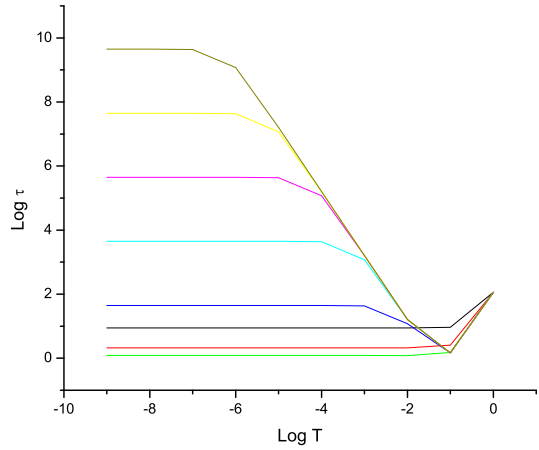, width=8cm}} \caption{The time
scales as function of temperature for a strange star with
$M=1.4M_\odot, R=10$km, ${\cal T}=1.5$ms for relative perturbation
$10^{-2}, 10^{-1}, 10^0, 10^{-3}, 10^{-4}, 10^{-5}, 10^{-6},
10^{-7}$ from bottom to top.}\label{mfig4}
\end{figure}

 Figure 4-7show the damping time as functions of temperature for some typical spinning strange star
 with periods  ${\cal T}=1.5$ms, 3ms, 15ms and Keplian limit. In
 comparison with harmonic oscillations, the  damping time scale of
 r-mode with period $\tau={2\over 3}{\cal T}=10^{-3}$s(figure\ref{mfig4})
 is only short to about 1s, but that of harmonic oscillations can be
 short to $10^{-3}$s(see figure 4 in Ref 17).
This implies that the simple radial pulsations can be
 damped more efficiently by the bulk viscosity. For a Keplian rotation
 star, the minimum time scale is about $10^{-1}$s.

 \section{CONCLUSION}

 Considering the self-consistent thermodynamics in density
 dependent quark mass model, we correct the error in\cite{21} and re-derive the bulk viscosity of SQM.
 The viscosity in DDQM is larger than MIT bag model. Consequences
 is that the r-modes of strange star will be damped more
 efficiently in DDQM implying the r-mode instability window of strange stars should be modified.

 \begin{figure}[h]
\centerline{\epsfig{file=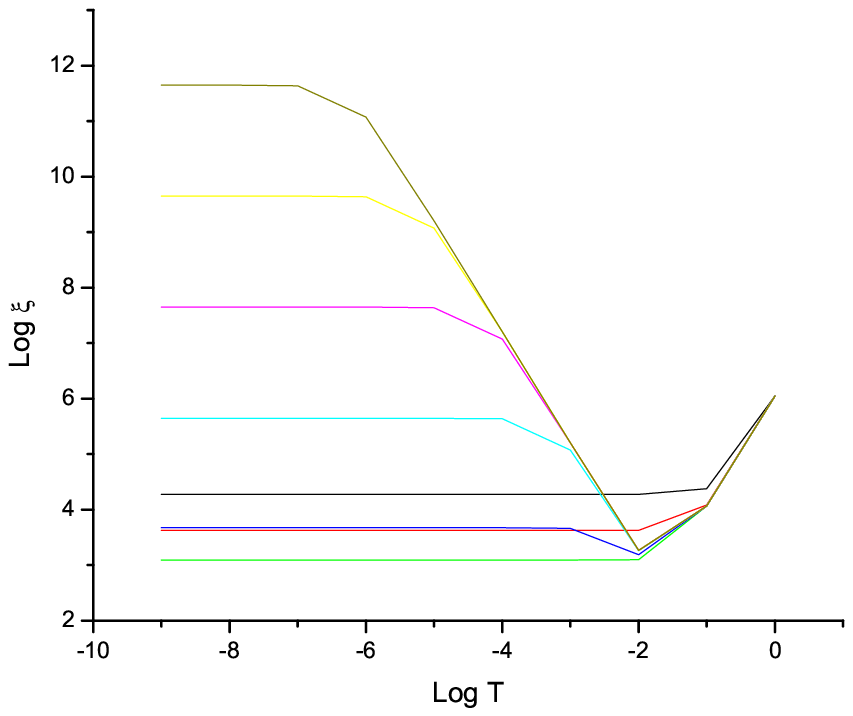, width=8cm}} \caption{ As Fig
4, but ${\cal T}=3$ms.}\label{mfig5}
\end{figure}
 \begin{figure}[h]
\centerline{\epsfig{file=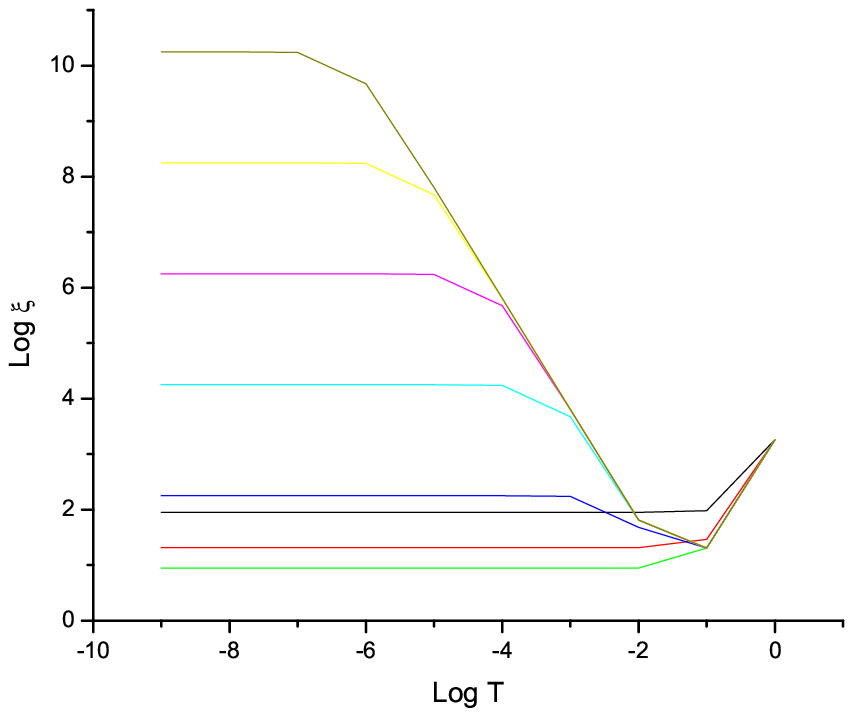, width=8cm}} \caption{ As Fig
4, but ${\cal T}=15$ms.}\label{mfig6}
\end{figure}
 \begin{figure}[h]
\centerline{\epsfig{file=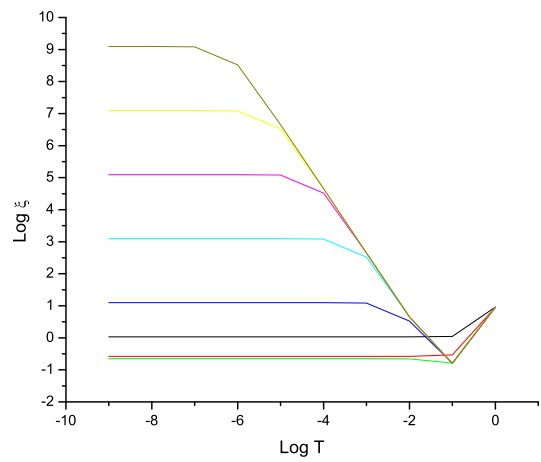, width=8cm}} \caption{ As Fig
4, but assuming a Keplian rotation star.}\label{mfig7}
\end{figure}
\end{document}